# Negative magnetoresistance due to conductivity fluctuations in films of the topological semimetal $Cd_3As_2$

Timo Schumann[1,a)], Manik Goyal[1], David A. Kealhofer[2], and Susanne Stemmer[1,b)]

[1]*Materials Department, University of California, Santa Barbara, California 93106-5050, USA*

[2]*Department of Physics, University of California, Santa Barbara, California 93106-9530, USA*

a) Email: schumann.timo@gmx.net

b) Email: stemmer@mrl.ucsb.edu

**Abstract**

Recently discovered Dirac and Weyl semimetals display unusual magnetoresistance phenomena, including a large, non-saturating, linear transverse magnetoresistance and a negative longitudinal magnetoresistance. The latter is often considered as evidence of fermions having a defined chirality. Classical mechanisms, due to disorder or non-uniform current injection, can however, also produce negative longitudinal magnetoresistance. Here, we report on magnetotransport measurements performed on epitaxial thin films of $Cd_3As_2$, a three-dimensional Dirac semimetal. Quasi-linear positive transverse magnetoresistance and negative longitudinal magnetoresistance are observed. By evaluating films of different thickness and by correlating the temperature dependence of the carrier density and mobility with the magnetoresistance characteristics, we demonstrate that both the quasi-linear positive and the negative magnetoresistance are caused by conductivity fluctuations. Chiral anomaly is not needed to explain the observed features.

Recent experimental realizations of three-dimensional topological semimetals have opened up numerous exciting opportunities to test quasiparticle behavior that mimics massless, relativistic Dirac or Weyl fermions. A signature of Weyl fermions is the "chiral anomaly", when charge flows from one Weyl node to one of opposite chirality under parallel electric and magnetic fields, thereby giving rise to an unusual *negative* longitudinal magnetoresistance (MR) [1, 2]. In Dirac semimetals the Dirac nodes split into Weyl nodes in the magnetic field and similar physics is expected. Recent observations [3-9] of longitudinal negative MR in Dirac and Weyl semimetals have therefore generated significant attention. While most studies consider the negative longitudinal MR to be evidence of the chiral anomaly, current jetting has been proposed as an alternative explanation [10, 11]. Current jetting arises under inhomogeneous current injection into high mobility materials that have a large conductivity anisotropy [12]. It leads to a strong preference of the current to flow in the direction of the magnetic field, resulting in a negative longitudinal MR [12]. Other studies claim to have ruled out current jetting caused by non-uniform current injection [7]. Negative longitudinal MR can also be due to conductivity fluctuations and has been reported for disordered semiconductors [13, 14]. A recent study [15] showed that mobility fluctuations are a likely origin of *linear, positive transverse MR* in $Cd_3As_2$, a three-dimensional Dirac semimetal [16-20].

Most studies that attribute the negative longitudinal MR to chiral anomaly have been carried out on bulk materials. Thin films offer several important advantages towards resolving the origin of the negative longitudinal MR in Dirac and Weyl semimetals. In particular, thin film device geometries are much less susceptible to current jetting. Furthermore, the effects of non-uniformities can be studied by varying the film thickness and growth conditions. In this work, we study the magnetotransport properties of epitaxial $Cd_3As_2$ thin films. We show that both the transverse positive MR and the negative longitudinal MR are due to conductivity fluctuations.

$Cd_3As_2$ films were grown by molecular beam epitaxy on relaxed, 180-nm-thick GaSb buffer layers on GaAs (111)B substrates (1° miscut towards the [$\overline{1}\overline{1}2$] direction), as described elsewhere [21]. The beam equivalent pressure was $2\times10^{-6}$ Torr, the substrate temperatures ranged between 150 °C and 170 °C, and the growth time was between 5 and 60 min (see Table I). Surface morphologies were investigated by optical, atomic force, and scanning electron microscopies. The film structure and alignment with the substrate was probed by x-ray diffraction (XRD) using Cu K-α radiation. The films were patterned into Hall bar structures with widths and lengths of 100 μm using standard optical lithography and $Ar^+$ ion milling. The thicknesses of the $Cd_3As_2$ layers were determined by scanning electron microscopy of the patterned Hall bar structures. Magneto-resistance measurements were carried out in a QuantumDesign Dynacool PPMS, at temperatures between 300 K and 2 K and magnetic fields up to 9 T. DC excitation currents of 10 μA were used in all measurements. Shubnikov-de Haas oscillations were detected in the thinnest films (Fig. S1 [22]).

Representative optical and atomic force micrographs of a $Cd_3As_2$ film are shown in Figs. 1(a,b). In most areas, the film is smooth with atomically stepped surfaces [Fig. 1 (b)]. The step height (~ 4 Å) matches the interplanar spacing of the $Cd_3As_2$ (112) planes, which form the surface [21]. The defects in the optical micrograph [Fig. 1 (a)] are likely areas that grew in a three-dimensional (island) growth mode. Their density was about $10^3 - 10^4$ $cm^{-2}$. Out-of-plane XRD confirmed single phase, epitaxial $Cd_3As_2$ [Fig. 1(c)]. High-resolution XRD showed Laue fringes from the GaSb buffer layer and for the thinnest $Cd_3As_2$ layers [arrows in Fig. 1 (d)]. The three-dimensional carrier densities determined from low field Hall measurements were similar for all films grown at 150 °C (see Table I) and ranged from ~ (0.7 – 1.8)×10$^{17}$ cm$^{-3}$ at 2 K to (5.0 – 8.0)×10$^{17}$ cm$^{-3}$ at 300 K. These carrier densities are lower than those of single crystals reported in the literature (typically > 10$^{18}$ cm$^{-3}$ [15, 17, 23]), indicating that the Fermi level is lower (i.e., closer to the Dirac node). The carrier density of the film grown at 170 ˚C was slightly higher, ~ 3.5×10$^{17}$ cm$^{-3}$ at 2 K (14.2×10$^{17}$ cm$^{-3}$ at 300 K). The phase diagram for Cd$_3$As$_2$ shows a line compound that opens up to a phase field towards the Cd-rich side, i.e. Cd$_3$As$_{2-\delta}$, at higher temperatures [24]. This leads to higher concentrations of As vacancies, which are the source of unintentional *n*-type doping in Cd$_3$As$_2$ [25]. The Hall mobility (Table I) increased with film thickness either due to a reduction in scattering from the interface with the buffer layer or annihilation of extended defects with increasing thickness.

Figure 2 shows the transverse (magnetic field **B** perpendicular to the electric field **E**) and longitudinal (**B** parallel to **E**) MR of the different samples, measured at 2 K. For **B** normal to the film plane [Fig. 2(a)], a positive, quasi-linear, non-saturating MR is observed, similar to many other Dirac and Weyl semimetals. A linear MR in Cd$_3$As$_2$ has been attributed to the lifting of the protection from backscattering under an applied magnetic field [23, 26] or due to the quantum magnetoresistance [27]. The latter condition (all electrons in the lowest Landau level) is not fulfilled here. Charge density and/or mobility fluctuations can also cause linear, non-saturating MR [14, 28-31]. In an inhomogeneous material, the direction of travel of the charge carriers deviates from the direction of the applied field, leading to a velocity component perpendicular to both **B** and **E**. A Hall signal is superimposed, leading to a contribution linear in **B**. Measurements with **B** in the film plane and perpendicular to the current are shown in Fig. 2(b). For thicker films, a positive, quasi-linear MR is observed. In contrast, for the thinner samples (<170 nm), the magnitude of the positive MR is reduced and resembles classical, ~ $B^2$ behavior.

With **B** in the film plane, carriers would need to travel along the film normal to generate a Hall contribution. In thin films, confinement makes this path unavailable and the linear contribution to the MR vanishes. Thus the film thickness dependence is a clear indication that the quasi-linear, positive transverse MR arises from a classical mechanism, in accordance with the conclusions in ref. [15], and is not a phenomenon reflecting the peculiar properties of Dirac fermions.

Negative longitudinal MR is observed in all samples [Fig 2(c)], followed by a positive MR at high **B**. Additionally, positive MR is observed in thicker films at low **B**. These features are strikingly similar to those reported for bulk, thin film, and nanowire $Cd_3As_2$ [6, 32-34], as well as other Dirac and Weyl semimetals [4, 5, 7, 8]. We also measured much thicker films, which have higher mobilities (see ref. [21]) and which also showed qualitatively similar behavior. We first discuss the positive longitudinal MR component. While a weak positive low-field MR in $Cd_3As_2$ can be attributed to weak antilocalization [8, 32, 35], the correction to the conductance is too strong and the temperature dependence too weak for weak localization (see [22]). A longitudinal positive MR of $Cd_3As_2$ has previously been observed and attributed to the two elliptical Fermi surfaces centered at the Dirac nodes, which are displaced away from the zone center [36]. This seems a reasonable explanation for the positive MR component observed here, which is $\sim B^2$ (see below).

We next address the strong negative component of the longitudinal MR. In thin film experiments the current is injected by large-area Ohmic contacts away from the measured Hall bar structures (>200 μm, see Fig. S4 [22]), so geometric current jetting, see ref. [10], is likely not the origin of the negative longitudinal MR. To further exclude geometrical current jetting, additional measurements with completely straight Hall bars were conducted, which show similar behavior (see Fig S5 [22]). To gain further insights, we use the following generic equation to describe the longitudinal conductivity at different temperatures:

$$\sigma(B) = \sigma_0 + S_p B^2 + \frac{1}{R_n + A_n B^2}. \tag{1}$$

Eq. (1) contains two terms that are quadratic in $B$, with one leading to a negative MR or positive magnetoconductance ($S_p$) and the other to a positive MR (consistent with prior observations [36], see above) or negative magnetoconductance ($R_n$ and $A_n$). Similar equations have been used to describe Dirac and Weyl semimetals in the literature [4, 5, 35], because theory predicts that the magnetoconductance due to the chiral anomaly in the classical (not quantum) regime is $\sim B^2$ [2]. Here, we use Eq. (1) because it adequately describes the magnetoconductance between 2 K and 300 K [see Fig. 3(a)] with a minimum number of terms. This allows us to evaluate the relative strengths of the positive and negative contributions as a function of temperature. Figures 3(c,d) show the extracted fit parameters as a function of temperature. $S_p$ increases with temperature, reaching a peak value at $\sim 100$ K, and subsequently decreases. In contrast, $R_n$ and $A_n$ first decrease with increasing temperature, reaching a minimum at $\sim 100$ K, and then increase.

Crucially, the MR behavior can be compared with the temperature behavior of charge carrier densities and mobilities [Fig. 3(b)]. The results in Fig. 3(b) were obtained by fitting the Hall resistance, $R_{xy}(B)$, which was non-linear at high **B**, to the standard two carrier Drude equation [22, 37]. The zero-field longitudinal resistance was set to $R_{xx}(0\text{ T}) = (n_1 e)^{-1} + (n_2 e)^{-1}$, where $n_1$ and $n_2$ (are the carrier densities and $e$ the elementary charge. The transverse magnetoresistance with **B** out of plane cannot be fit to the standard two carrier model as the linear component (discussed above) is not described by it. Both carriers are n-type. Carrier 1 has higher mobility ($\sim 1.2$ m$^2$/Vs at 300 K and $\sim 1.45$ m$^2$/Vs at 2 K) and densities ($5\times 10^{13}$ cm$^{-2}$ at 300 K and $1.1\times10^{13}$ cm$^{-2}$ at 2 K) than carrier 2. The presence of two carrier types is consistent with the electronic structure of Cd$_3$As$_2$, which exhibits a conduction band valley at the zone center that is only slightly higher in energy than the Dirac nodes [16, 19].

Most importantly, the temperature dependence of the carrier densities exhibits a crossover behavior. The carrier densities are nearly independent of temperature at low temperatures but increase with temperature above ~ 100 K. The density of carrier 1 in particular shows a steep increase. The crossover coincides with the maximum of $S_p$. This correlation supports the notion that the negative longitudinal MR is caused by conductivity fluctuations. In particular, spatial fluctuations in the carrier density and/or mobility can lead to inhomogeneous equipotential lines and a negative longitudinal MR [13, 38]. Conductivity fluctuations result in current paths that are not parallel to the applied electric field and thus are similar to current jetting. The effects of conductivity fluctuations are strongest in crossover regimes, such as around 100 K, where inhomogeneities lead to large relative mobility/charge carrier density differences [13]. At lower temperatures, as well as at higher temperatures, carrier 1 dominates the transport, leading to reduced relative conductivity fluctuations and therefore reduced positive magnetoconductance (negative MR). The films contain twin defects [21] and likely also As vacancies, which may form clusters [20]. The importance of the latter is corroborated by the fact that the negative longitudinal MR is strongest in the film grown at higher temperatures, which possesses the highest carrier density and hence As vacancy concentration. Simulations of the longitudinal MR due to conductivity fluctuations have been carried out for simple model geometries in the literature [31, 38], confirming that they are a cause of negative longitudinal MR. They lead to results that are remarkably similar to those observed here [38]. Quantitative simulation of the MR for the films would require a detailed model of the spatial distribution and nature of the inhomogeneities. Interestingly, other topological Dirac materials for which the negative longitudinal MR has been observed also show a strongly temperature-dependent Hall coefficient (carrier density) [5]. Further studies are clearly needed to understand the temperature-dependent carrier densities and mobilities in all of these materials. The thickness dependence of the longitudinal negative MR further corroborates the interpretation that it is due to conductivity fluctuations. Like the positive linear transverse magnetoresistance, it is less pronounced in thinner films [see Fig. 2(c)]. Current distortions require a path normal to the film thickness, which gets increasingly cut off in thinner films [38].

To summarize, the magnetotransport properties of $Cd_3As_2$ thin films are remarkably similar to those found in other Dirac and Weyl semimetals, exhibiting both a positive linear transverse MR and the negative longitudinal MR. The film thickness dependence and the correlation with a crossover in the temperature dependence of the carrier density establish that the negative longitudinal MR is due to charge carrier density and/or mobility fluctuations causing distortions in the current paths. Although the results do not rule out the possibility of observing the chiral anomaly in magnetoresistance measurements, they clearly show that the chiral anomaly does not need to be invoked to explain any of the observed features. Geometrical avoidance of current jetting is not sufficient evidence that the negative longitudinal MR is due to the chiral anomaly, if the material contains disorder.

**Acknowledgements**

The authors gratefully acknowledge support through the Vannevar Bush Faculty Fellowship program by the U.S. Department of Defense (grant no. N00014-16-1-2814). This work made use of the MRL Shared Experimental Facilities, which are supported by the MRSEC Program of the U.S. National Science Foundation under Award No. DMR 1121053.

# References

[1] H. B. Nielsen, and M. Ninomiya, Phys. Lett. B **130**, 389 (1983).

[2] D. T. Son, and B. Z. Spivak, Phys. Rev. B **88**, 104412 (2013).

[3] Q. Li, D. E. Kharzeev, C. Zhang, Y. Huang, I. Pletikosi, A. V. Fedorov, R. D. Zhong, J. A. Schneeloch, G. D. Gu, and T. Valla, Nat. Phys. **12**, 550 (2016).

[4] X. Huang, L. Zhao, Y. Long, P. Wang, D. Chen, Z. Yang, H. Liang, M. Xue, H. Weng, Z. Fang, X. Dai, and G. Chen, Phys. Rev. X **5**, 031023 (2015).

[5] J. Xiong, S. K. Kushwaha, T. Liang, J. W. Krizan, M. Hirschberger, W. Wang, R. J. Cava, and N. P. Ong, Science **350**, 413 (2015).

[6] C.-Z. Li, L.-X. Wang, H. Liu, J. Wang, Z.-M. Liao, and D.-P. Yu, Nat. Comm. **6**, 10137 (2015).

[7] M. Hirschberger, S. Kushwaha, Z. Wang, Q. Gibson, S. Liang, C. A. Belvin, B. A. Bernevig, R. J. Cava, and N. P. Ong, Nat. Mater. **15**, 1161 (2016).

[8] C. Zhang, S.-Y. Xu, I. Belopolski, Z. Yuan, Z. Lin, B. Tong, N. Alidoust, C.-C. Lee, S.-M. Huang, H. Lin, M. Neupane, D. S. Sanchez, H. Zheng, G. Bian, J. Wang, C. Zhang, T. Neupert, M. Z. Hasan, and S. Jia, Nat. Comm. **7**, 10735 (2016).

[9] A. C. Niemann, J. Gooth, S.-C. Wu, S. Ba ßler, P. Sergelius, R. Hu hne, B. Rellinghaus, C. Shekhar, V. Su ß, M. Schmidt, C. Felser, B. Yan, and K. Nielsch, Sci. Rep. **7**, 43394 (2017).

[10] R. D. dos Reis, M. O. Ajeesh, N. Kumar, F. Arnold, C. Shekhar, M. Naumann, M. Schmidt, M. Nicklas, and E. Hassinger, New J. Phys. **18**, 085006 (2016).

[11]  F. Arnold, C. Shekhar, S.-C. Wu, Y. Sun, R. D. d. Reis, N. Kumar, M. Naumann, M. O. Ajeesh, M. Schmidt, A. G. Grushin, J. H. Bardarson, M. Baenitz, D. Sokolov, H. Borrmann, M. Nicklas, C. Felser, E. Hassinger, and B. Yan, Nat. Comm. **7**, 11615 (2016).

[12]  A. B. Pippard, *Magnetoresistance in Metals* (Cambridge University Press, Cambridge, 1989).

[13]  J. Hu, T. F. Rosenbaum, and J. B. Betts, Phys. Rev. Lett. **95**, 186603 (2005).

[14]  T. Khouri, U. Zeitler, C. Reichl, W. Wegscheider, N. E. Hussey, S. Wiedmann, and J. C. Maan, Phys. Rev. Lett. **117**, 256601 (2016).

[15]  A. Narayanan, M. D. Watson, S. F. Blake, N. Bruyant, L. Drigo, Y. L. Chen, D. Prabhakaran, B. Yan, C. Felser, T. Kong, P. C. Canfield, and A. I. Coldea, Phys. Rev. Lett. **114**, 117201 (2015).

[16]  Z. J. Wang, H. M. Weng, Q. S. Wu, X. Dai, and Z. Fang, Phys. Rev. B **88**, 125427 (2013).

[17]  M. Neupane, S. Y. Xu, R. Sankar, N. Alidoust, G. Bian, C. Liu, I. Belopolski, T. R. Chang, H. T. Jeng, H. Lin, A. Bansil, F. Chou, and M. Z. Hasan, Nat. Comm. **5**, 3786 (2014).

[18]  S. Borisenko, Q. Gibson, D. Evtushinsky, V. Zabolotnyy, B. Buchner, and R. J. Cava, Phys. Rev. Lett. **113**, 165109 (2014).

[19]  Z. K. Liu, J. Jiang, B. Zhou, Z. J. Wang, Y. Zhang, H. M. Weng, D. Prabhakaran, S.-K. Mo, H. Peng, P. Dudin, T. Kim, M. Hoesch, Z. Fang, X. Dai, Z. X. Shen, D. L. Feng, Z. Hussain, and Y. L. Chen, Nat. Mater. **13**, 677 (2014).

[20]  S. Jeon, B. B. Zhou, A. Gyenis, B. E. Feldman, I. Kimchi, A. C. Potter, Q. D. Gibson, R. J. Cava, A. Vishwanath, and A. Yazdani, Nat. Mater. **13**, 851 (2014).

[21]  T. Schumann, M. Goyal, H. Kim, and S. Stemmer, APL Mater. **4**, 126110 (2016).

[22] See Supplemental Material [link to be inserted by publisher] for Shubnikov-de Haas oscillations, weak antilocalization fits, two-carrier fits of $R_{xy}$, a micrograph of the Hall bar structure, and results from a Hall bar structure that has a slightly different geometry.

[23] T. Liang, Q. Gibson, M. N. Ali, M. H. Liu, R. J. Cava, and N. P. Ong, Nat. Mater. **14**, 280 (2015).

[24] H. Baker, *Alloy Phase Diagrams* (ASM International, Materials Park (Ohio), 1992), Vol. 3.

[25] D. P. Spitzer, G. A. Castellion, and G. Haacke, J. Appl. Phys. **37**, 3795 (1966).

[26] J. Feng, Y. Pang, D. Wu, Z. Wang, H. Weng, J. Li, X. Dai, Z. Fang, Y. Shi, and L. Lu, Phys. Rev. B **92**, 081306(R) (2015).

[27] A. A. Abrikosov, Phys. Rev. B **58**, 2788 (1998).

[28] M. M. Parish, and P. B. Littlewood, Nature **426**, 162 (2003).

[29] N. V. Kozlova, N. Mori, O. Makarovsky, L. Eaves, Q. D. Zhuang, A. Krier, and A. Patane, Nat. Commun. **3**, 1097 (2012).

[30] J. Hu, and T. F. Rosenbaum, Nat. Mater. **7**, 697 (2008).

[31] F. Kisslinger, C. Ott, and H. B. Weber, Phys. Rev. B **95**, 024204 (2017).

[32] B. Zhao, P. H. Cheng, H. Y. Pan, S. Zhang, B. G. Wang, G. H. Wang, F. X. Xiu, and F. Q. Song, Sci. Rep. **6**, 22377 (2016).

[33] H. Li, H. T. He, H. Z. Lu, H. C. Zhang, H. C. Liu, R. Ma, Z. Y. Fan, S. Q. Shen, and J. N. Wang, Nat. Comm. **7**, 10301 (2016).

[34] J. Xiong, S. K. Kushwaha, T. Liang, J. W. Krizan, W. Wang, R. J. Cava, and N. P. Ong, arXiv:1503.08179 [cond-mat.str-el] (2015).

[35] J. Hu, J. Y. Liu, D. Graf, S. M. A. Radmanesh, D. J. Adams, A. Chuang, Y. Wang, I. Chiorescu, J. Wei, L. Spinu, and Z. Q. Mao, Sci. Rep. **6**, 18674 (2016).

[36] M. Iwami, H. Matsunami, and T. Tanaka, J. Phys. Soc. Jpn. **31**, 768 (1971).

[37] D. C. Look, *Electrical Characterization of GaAs Materials and Devices* (Wiley, New York, 1989).

[38] J. S. Hu, M. M. Parish, and T. F. Rosenbaum, Phys. Rev. B **75**, 214203 (2007).

**Table I:** Sample parameters of the $Cd_3As_2$ thin films. $n_{2D}$ is the sheet carrier density, $n_{3D}$ the three-dimensional carrier density and µ is the Hall mobility.

| | Growth temperature | Growth time (min) | Film thickness (nm) | $n_{2D}$ ($10^{12}$ cm$^{-2}$) | | $n_{3D}$ ($10^{17}$ cm$^{-3}$) | | µ (cm$^2$/Vs) | |
|---|---|---|---|---|---|---|---|---|---|
| | | | | 2 K | 300 K | 2 K | 300 K | 2 K | 300 K |
| o10 | 150 °C | 5  | 85  | 1.46 | 4.25  | 1.7 | 5.0  | 6700  | 4550 |
| o12 | 150 °C | 10 | 120 | 2.17 | 7.42  | 1.8 | 6.2  | 7900  | 6400 |
| o19 | 150 °C | 20 | 170 | 1.43 | 13.6  | 0.8 | 8.0  | 5250  | 8950 |
| o21 | 150 °C | 40 | 340 | 3.76 | 24.1  | 1.1 | 7.1  | 11250 | 11950 |
| o22 | 170 °C | 60 | 370 | 13.0 | 52.5  | 3.5 | 14.2 | 13650 | 12050 |

**Figure Captions**

**Figure 1:** (a) Optical micrograph of a $Cd_3As_2$ film patterned into a Hall bar. (b) Atomic force micrograph of the film surface. (c, d) XRD scans of a $Cd_3As_2$/GaSb/GaAs heterostructure. The arrows in (d) indicate the Laue fringes.

**Figure 2:** (a,b) Transverse (**B** perpendicular to **E**) magnetoresistance and (c) longitudinal (**B** in the plane and parallel to **E**) magnetoresistance obtained from different samples and measured at 2 K. The labels in (a.b) indicate the orientation of **B**: oop = out of plane and ip = in-plane. The samples are labeled in order of increasing thickness from sample o10 (85 nm) to o22 (370 nm), see Table I.

**Figure 3:** (a) Longitudinal conductivity for the $Cd_3As_2$ film grown at 170 ˚C for 60 min (sample number o22) as a function of temperature. The dotted lines are fits to the data, using Eq. (1), with the results shown in (c,d). (b) Carrier density and mobility extracted by fitting the Hall data to 9 T to a standard two carrier equation.

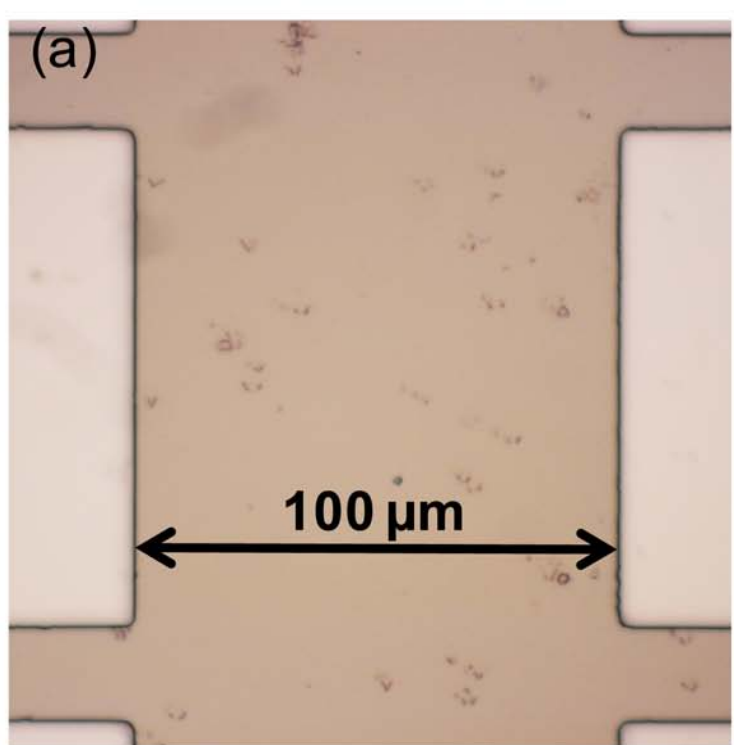
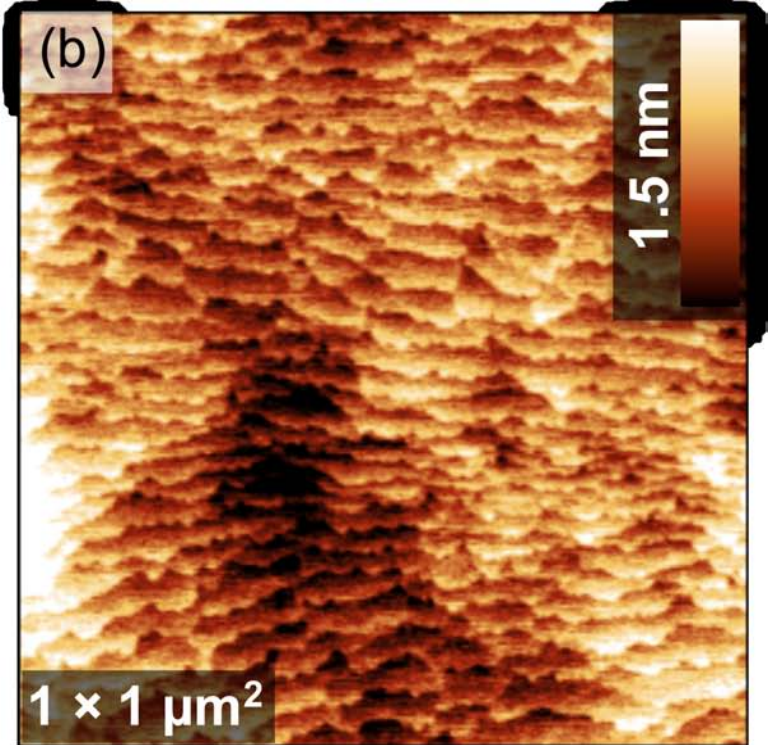
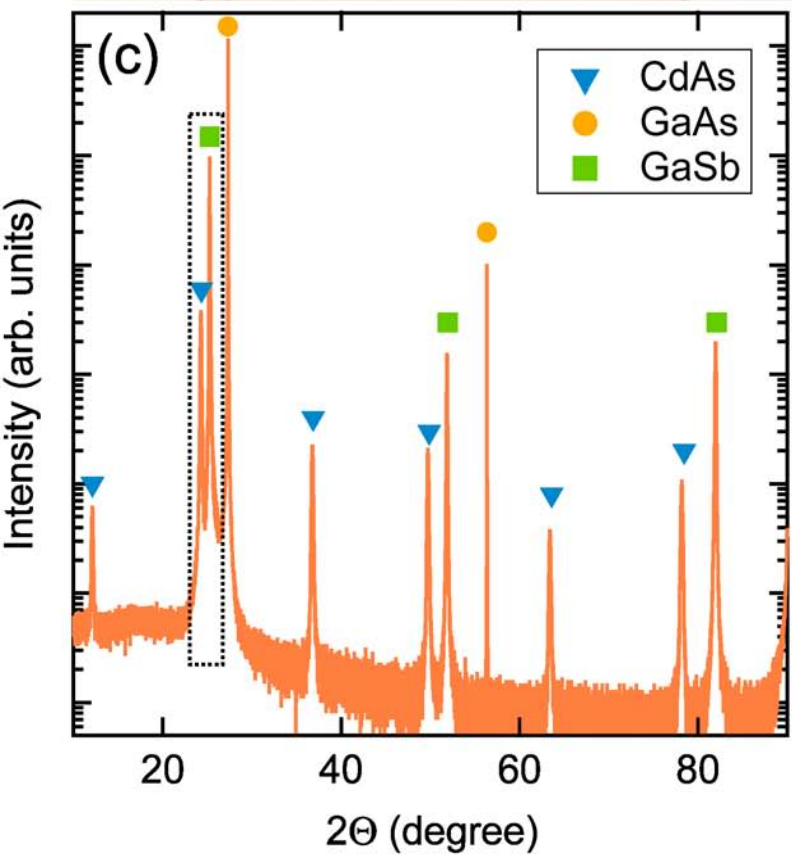
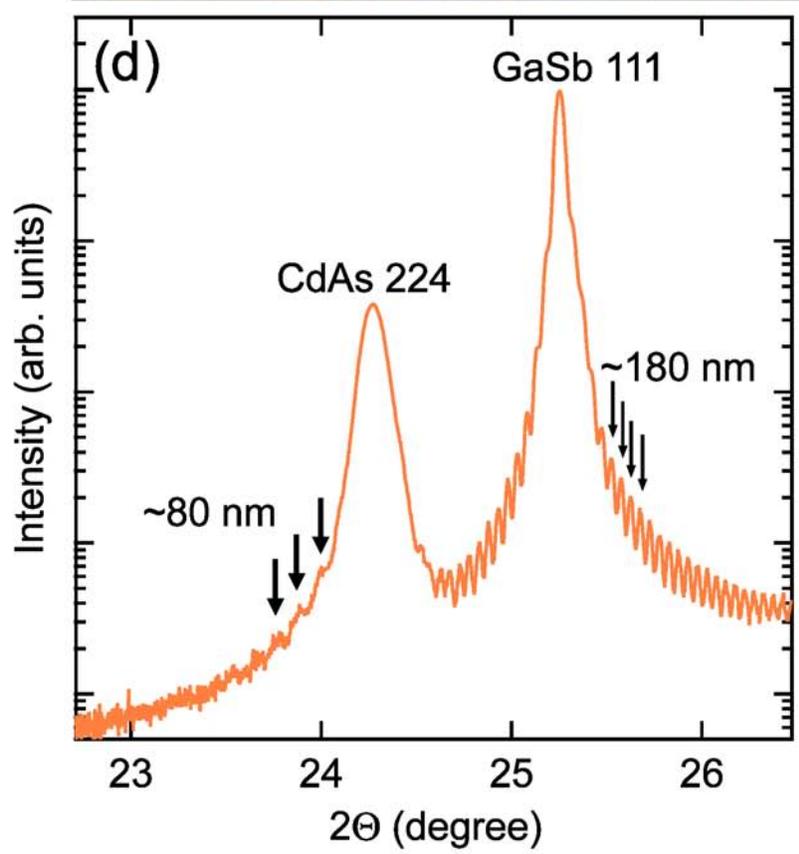

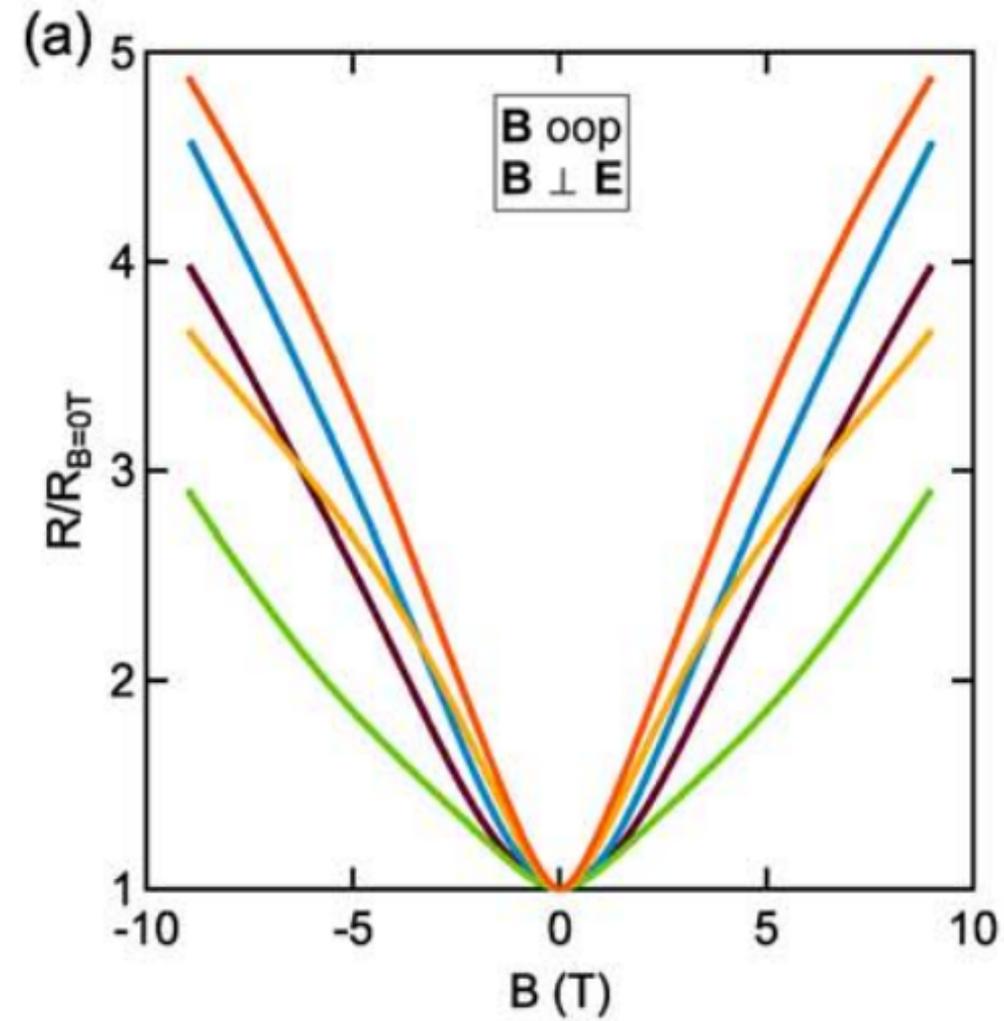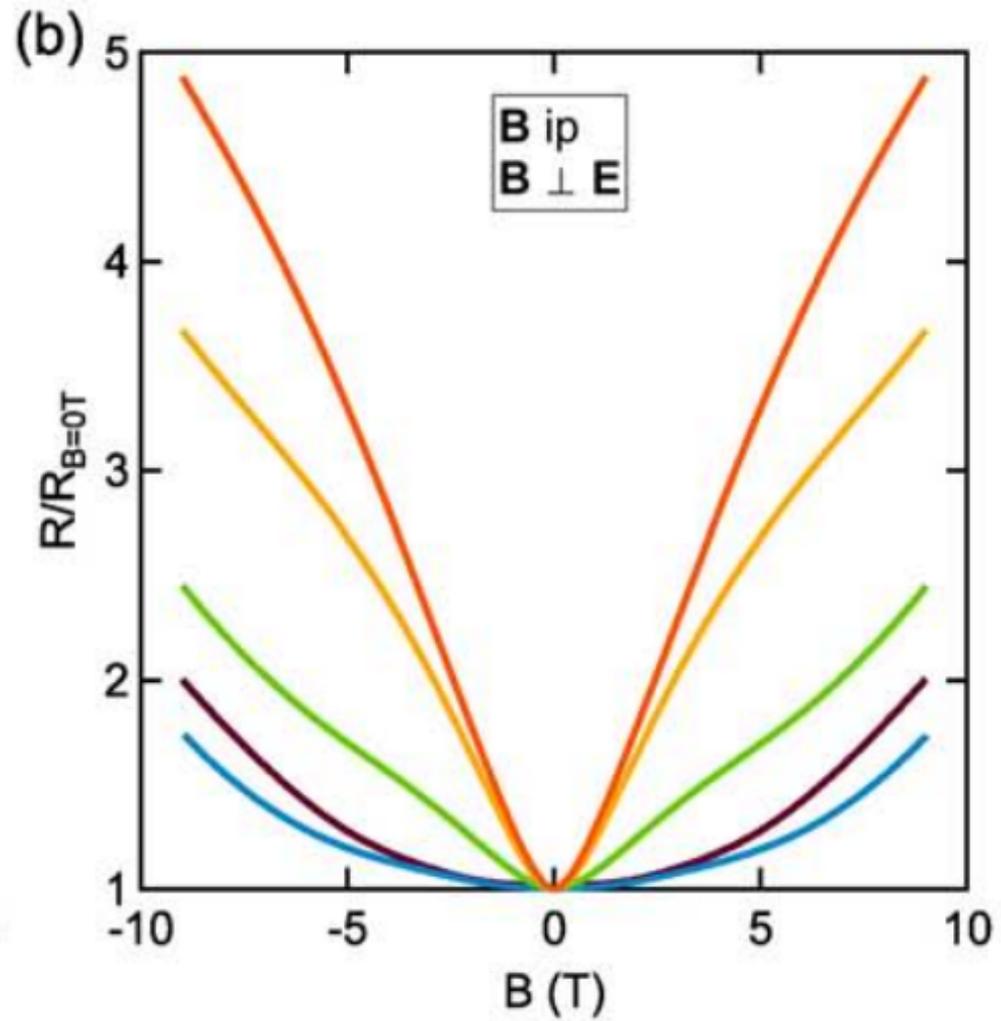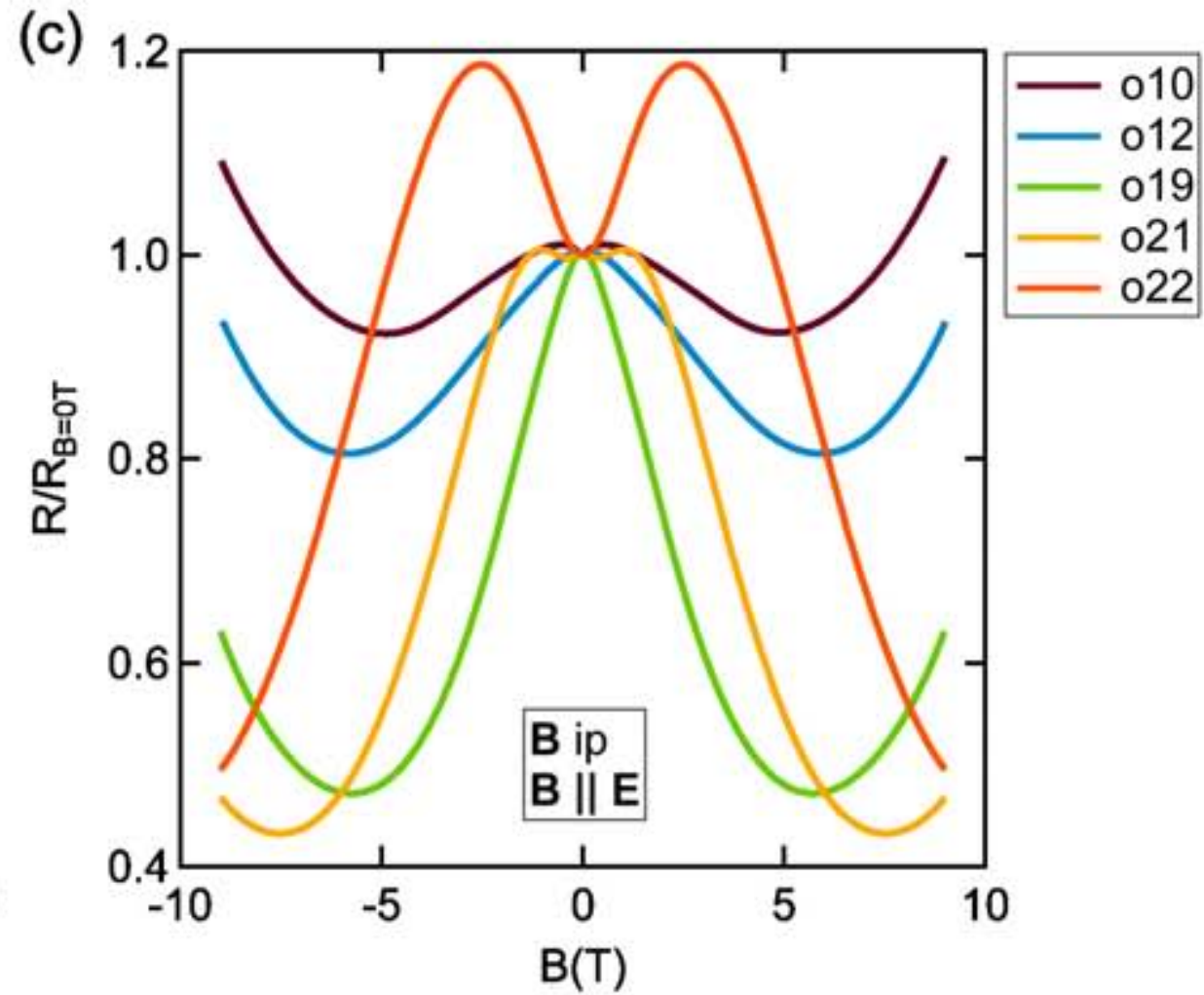

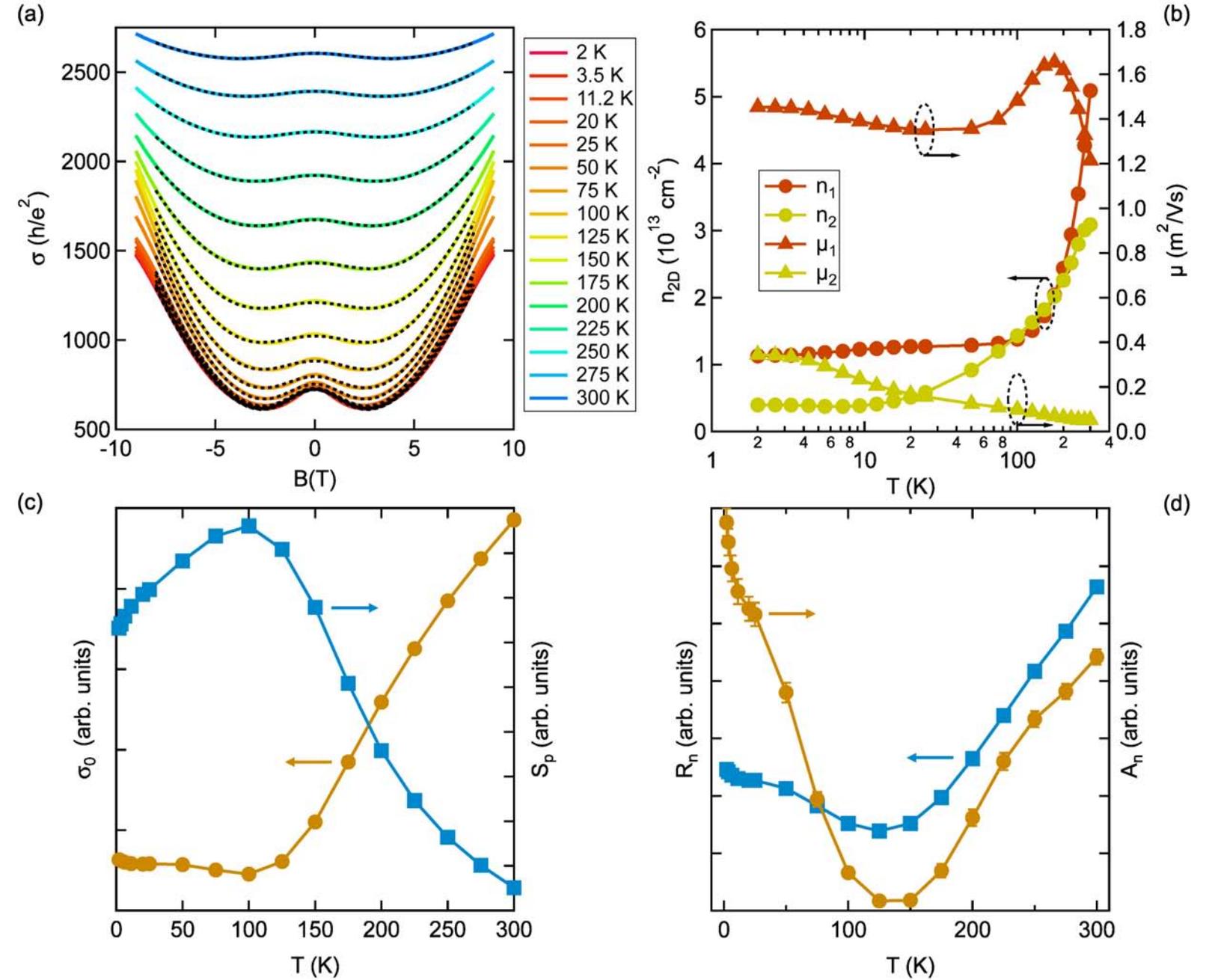

# Supplemental Information:
## Negative magnetoresistance due to conductivity fluctuations in films of the topological semimetal $Cd_3As_2$

Timo Schumann, Manik Goyal, David A. Kealhofer, and Susanne Stemmer

*Materials Department, University of California, Santa Barbara, California 93106-5050, USA*

**Magnetoresistance Oscillations**

Shubnikov-de Haas oscillations were observed in the transverse magnetoresistance obtained from sample o10, as shown in Fig. S1(a). The second derivative and its Fourier transform are shown in Figs. S1(b) and (c), respectively. The presence of the oscillations confirms the high quality of the films. A frequency of ~12 T is extracted, corresponding to a carrier concentration of ~ $6 \times 10^{11}$ cm$^{-2}$. This is lower than the carrier concentration extracted from the Hall measurements (see main text) and shows that only a fraction of the carriers with sufficient mobility give rise to the oscillations.

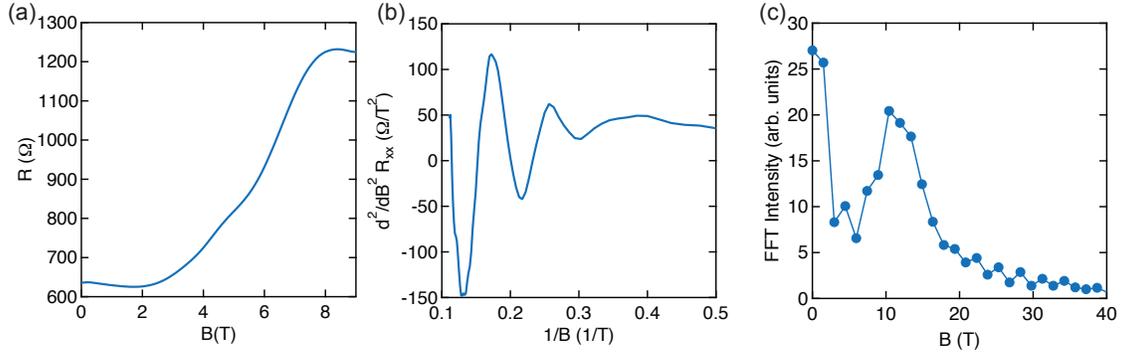

**Figure S1:** a) Transverse magnetoresistance. (b) Second derivative of (a), showing Shubnikov-de Haas oscillations. (c) Fourier transform of (b).

**Weak antilocalization**

Weak antilocalization (WAL) is observed both in transverse and longitudinal magnetoresistance of the thinnest samples, similar to $Cd_3As_2$ thin films in the literature [1]. In the strong spin-orbit coupling limit WAL is described by [2]:

$$\Delta\sigma(B) = \alpha \frac{e^2}{2\pi^2\hbar}\left(ln\left(\frac{B_\phi}{B}\right) - \psi\left(\frac{1}{2} + \frac{B_\phi}{B}\right)\right), \quad (S1)$$

where $\alpha$ is -1 for weak localization and +½ for weak anti-localization, $B_\phi$ is the phase coherence characteristic field and $\psi$ the Digamma function. Here, $\sigma$ is defined as $1/R_{xx}$, which is a good approximation since $R_{xx} \gg R_{xy}$ in the $B$ field range. Measurements at different temperatures and the fits to Eq. (S1) are shown in Fig. S2(a). The correction to the MR is well described by WAL. $B_\phi$ is connected to the phase coherence length $l_\phi$, via $l_\phi = (\hbar/4eB_\phi)^{1/2}$. The extracted parameters are shown in Fig. S2(b). At low temperatures, α is close to ½, as expected for WAL. The coherence length is about 140 nm at 2 K and reduces with increasing temperature. Note that the correction to the conductivity is less than one conduction quantum ($h/e^2$), and therefore several orders of magnitude weaker than the large positive longitudinal MR at low fields seen in Fig. 2(c) and discussed in the main text.

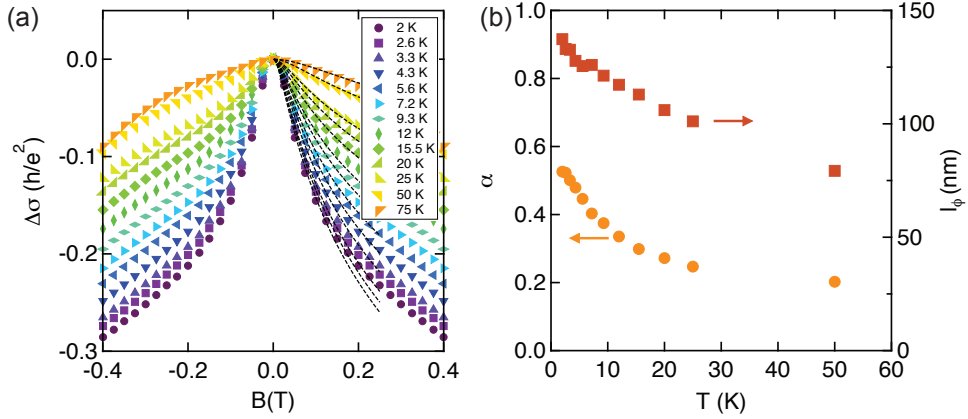

**Figure S2:** (a) Magnetoconductance and WAL fits at low fields of sample o10. (b) Extracted fit parameters.

**Two-carrier transport**

Hall measurement performed on sample o22 with B out-of-plane, together with fitted curves for two-carrier transport for selected temperatures are shown in Fig. S3.
The data can be described by:

$$R_{xy} = \frac{B}{e} \frac{(\mu_1^2 n_1 + \mu_1^2 n_1) + (\mu_1 \mu_2 B)^2 (n_1 + n_2)}{(\mu_1 n_1 + \mu_2 n_2)^2 + (\mu_1 \mu_2 B)^2 (n_1 + n_2)^2}, \quad (1)$$

where $e$ is the elementary electric charge, $B$ the magnetic field strength, and $n_i$ and $\mu_i$ ($i = 1,2$) charge carrier density and mobility for the two carriers, respectively.

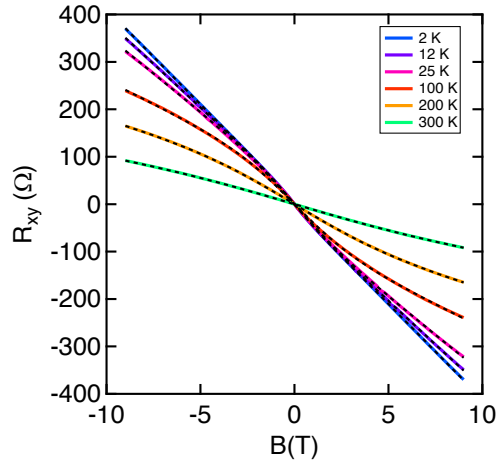

**Figure S3:** $R_{xy}$ measured on sample o22 with **B** out-of plane for selected temperatures. The fitted curves are shown by dotted lines.

**Micrograph of the Hall bar structure**

Figure S4 shows an optical micrograph of the device used for the measurements reported in the main text.

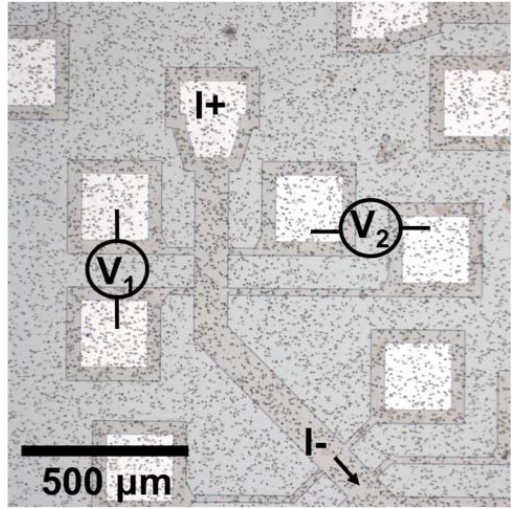

**Figure S4:** Large-area optical micrograph of a patterned Hall bar structure with the measurement pads indicated.

**Excluding geometrical current jetting**

As can be seen from Fig. S4, the current was injected in a large distance (>200 µm) from the measured Hall bar structure. There are kinks in the Hall bar away from the measurement terminals, which allows for convenient investigation of different geometries (**B** in-plane and parallel as well as perpendicular to the current). To exclude influence of these kinks, such as a current gradient, a simpler, straight Hall bar structure was defined on a different film [Fig. S5(a)]. Measurements conducted on both sides of the Hall bar structure show similar behavior [S5(c,d)] and the extracted fitting parameters [S5 (e,f)] are similar to those shown in the main manuscript.

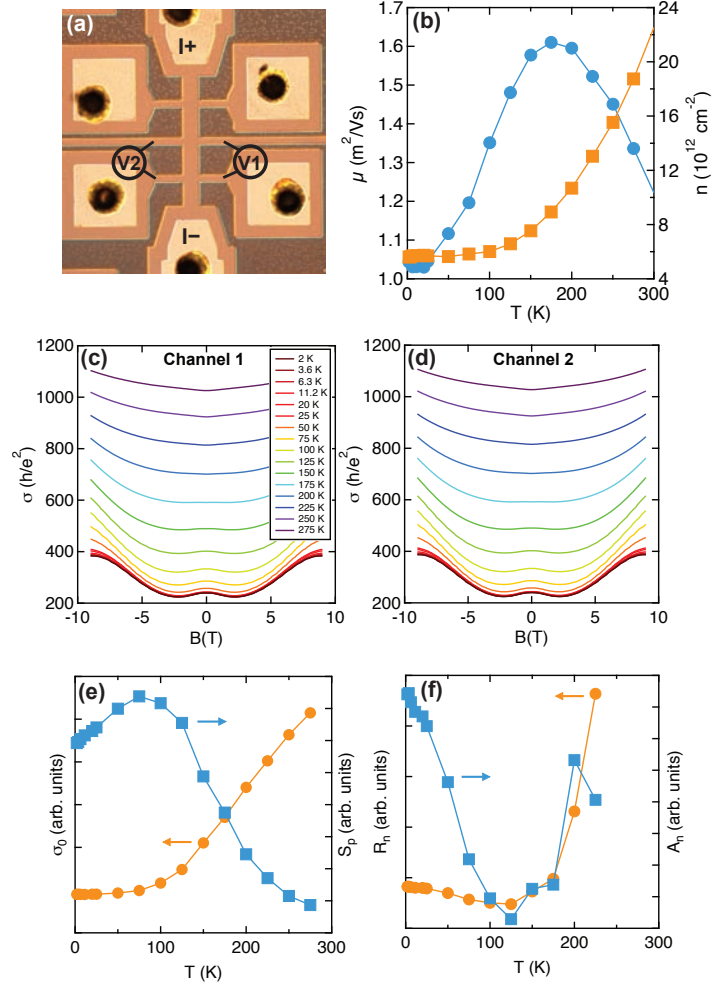

**Figure S5**: (a) Optical micrograph of a straight Hall bar structure. (b) Extracted carrier density and Hall mobility. (c,d) Longitudinal conductivity (**B**||**E**) measured on the two opposite sides of the Hall bar structure. (e,f) Extracted fitting parameters, see main manuscript for details.

## References

[1] B. Zhao, P. H. Cheng, H. Y. Pan, S. Zhang, B. G. Wang, G. H. Wang, F. X. Xiu, and F. Q. Song, *Weak antilocalization in $Cd_3As_2$ thin films*, Sci. Rep. **6**, 22377 (2016).

[2] S. Hikami, A. I. Larkin, and Y. Nagaoka, *Spin-Orbit Interaction and Magnetoresistance in the Two Dimensional Random System*, Prog. Theor. Phys. **63**, 707-710 (1980).